\documentstyle [twocolumn,prb,aps] {revtex}
\begin{document}
\draft
\begin{title}
{\bf On mesoscopic forces and quantized conductance in model metallic
nanowires}
\end{title} 
\author{Constantine Yannouleas and Uzi Landman}
\address{
School of Physics, Georgia Institute of Technology,
Atlanta, Georgia 30332-0430 }
\date{LETTER in J. Phys. Chem. B {\it 101\/}, 5780 [24 July 1997]}
\maketitle
\begin{abstract}
Energetics and conductance in jellium modelled nanowires are investigated
using the local-density-functional-based shell correction method. In analogy
with studies of other finite-size fermion systems , e.g., simple-metal
clusters or $^3$He clusters, we find that the energetics of the wire as a
function of its radius (transverse reduced dimension) leads to formation
of self-selecting magic wire configurations (MWC's, i.e., discrete
sequence of wire radii with enhanced stability), originating from 
quantization of the electronic spectrum, namely formation of subbands which
are the analogs of electronic shells in clusters. These variations in the
energy result in oscillations in the force required to affect a transition from
one MWC of the nanowire to another, and are correlated directly with step-wise 
variations of the quantized conductance of the nanowire in units of
$2 e^2/h$.
\end{abstract}

\narrowtext

\section{Introduction}

Identification and understanding of the physical origins and systematics
underlying the variations of materials properties with size, form of 
aggregation, and dimensionality are some of the main challenges in modern
materials research, of ever increasing importance  in the face of the
accelerated trend toward miniaturization of electronic and mechanical
devices. While for over two decades studies of size-evolutionary patterns
of materials have focused on atomic and molecular clusters 
\cite{haber,heer,yann1} in beams or embedded
in inert matrices, more recent efforts concentrated on preparation,
characterization, and understanding of finite solid-sate structures. These
include nanometer-scale metal and semiconductor 
nanocrystals, \cite{whet,lued,aliv}
surface-supported structures and quantum dots, \cite{avou}
and nanoscale junctions or wires.$^{8-27}$

Interestingly, it has emerged that concepts and methodologies developed in the
context of isolated gas-phase clusters and atomic nuclei are often most useful
for investigations of finite-size solid-state structures. In particular, it has
been shown most recently \cite{land4,barn} 
through first-principles molecular dynamics
simulations that as metallic (sodium) nanowires are stretched to just a few
atoms in diameter, the reduced dimensions, increased surface-to-volume ratio, 
and impoverished atomic environment, lead to formation of structures, made of
the metal atoms in the neck, which can be described in terms of those 
observed in small gas-phase sodium clusters; hence they were termed as 
supported {\it cluster-derived structures (cds)\/}. The above prediction of
the occurrence of ``magic-number'' cds's in nanowires, due to characteristics
of electronic cohesion and atomic bonding in such structures of reduced
dimensions, are directly correlated with the energetics of metal clusters,
where magic-number sequences of clusters sizes, shapes and structural motifs 
due to electronic and/or geometric shell effects, have been long predicted and 
observed. \cite{heer,yann1,mart}

Furthermore, these results lead one directly to conclude that other properties
of nanowires, derived from their energetics, may be described using 
methodologies developed previously in the context of clusters. Indeed, in this
paper, we show that certain aspects of the mechanical response (i.e.,
elongation force) and electronic transport (e.g., quantized conductance) in
metallic nanowires can be analyzed using the local-density-approximation
(LDA) -based shell correction method (SCM), developed and applied previously
in studies of metal clusters. \cite{yann1,yann2}
Specifically, we show that in a jellium-modelled,
volume-conserving nanowire, variations of the total energy (particularly 
terms associated with electronic subband corrections) upon elongation of the
wire lead to {\it self-selection\/} of a sequence of stable ``magic''
wire configurations (MWC's, specified in our model by a sequence of the wire's
radii), with the force required to elongate the wire from one configuration to 
the next exhibiting an oscillatory behavior. Moreover, we show that due to the
quantized nature of electronic states in such wires, the electronic conductance
varies in a quantized step-wise manner (in units of the conductance quantum
$g_0=2e^2/h$), correlated with the transitions between MWC's and the 
above-mentioned force oscillations.

Prior to introducing the model studied in this paper, it is appropriate
to briefly review certain previous theoretical and experimental investigations,
which form the background and motivation for this study. Atomistic
descriptions, based on realistic interatomic interactions, and/or 
first-principles modelling and simulations played an essential role in
discovering the formation of nanowires, and in predicting and elucidating
the microscopic mechanisms underlying their mechanical, spectral, electronic
and transport properties. 

Formation and mechanical properties of 
interfacial junctions (in the form of 
crystalline nanowires) have been predicted through
early molecular-dynamics simulations, \cite{land1} where the materials 
(gold) were
modelled using semiempirical embedded-atom potentials. In these studies it has
been shown that separation of the contact between materials leads to 
generation of a connective junction which elongates and narrows through
a sequence of structural instabilities; at the early stages, elongation
of the junction involves multiple slip events, while at the later stages,
when the lateral dimension of the wire necks down to a diameter of about 15 
\AA, further elongation involves a succession of stress accumulation and
fast relief stages associated with a sequence of order-disorder structural
transformations localized to the neck region. \cite{land1,land2,land3}
These structural evolution
patterns have been shown through the simulations to be portrayed in
oscillations of the force required to elongate the wire, with a 
period approximately equal to the interlayer spacing. 
In addition, the ``sawtoothed'' character of the predicted force 
oscillations [see Fig.\ 3(b) in Ref.\ \onlinecite{land1} and
Fig.\ 3 in Ref.\ \onlinecite{land2}] reflects the stress accumulation and
relief stages of the elongation mechanism.
Moreover, the critical resolved yield stress of
gold nanowires has been predicted \cite{land1,land2}
to be $\sim$ 4GPa, which is 
over an order of magnitude larger than that of the bulk,
and is comparable to the theoretical value for Au (1.5 GPa) in the
absence of dislocations.

These predictions, as well as anticipated electronic conductance properties,
\cite{land1,boga1}
have been corroborated in a number of experiments using 
scanning tunneling and force microscopy, 
\cite{land1,pasc1,oles,pasc2,smith,rubi,stal} 
break junctions, \cite{krans} 
and pin-plate techniques \cite{costa,land2}
at ambient environments, as well as under 
ultrahigh vacuum and/or cryogenic conditions. Particularly, pertinent
to our current study are experimental observations of the oscillatory
behavior of the elongation forces and the correlations between the changes in 
the conductance and the force oscillations; see especially the simultaneous
measurements of force and conductance in gold nanowires in Ref.\ 
\onlinecite{rubi},
where in addition the predicted ``ideal'' value of the critical yield
stress has also been measured (see also Ref.\ \onlinecite{stal}).

The jellium-based model introduced in this paper,
which by construction is devoid of atomic crystallographic
structure, does not address issues pertaining to nanowire formation methods,
atomistic configurations, and mechnanical response modes [e.g., 
plastic deformation mechanisms, interplanar slip, ordering and disordering 
mechanisms (see 
detailed descriptions in Refs.\ \onlinecite{land1,land2}
and \onlinecite{land3}, 
and a discussion of conductance dips in Ref.\ \onlinecite{pasc2}),
defects, mechanichal reversibility, \cite{rubi,land2}
and roughening of the wires's morphology 
during elongation \cite{land3}], nor does it consider the effects of the above 
on the electron spectrum, transport properties, and dynamics \cite{barn}
Nevertheless, as shown below, the model offers a useful framework for linking 
investigations of solid-state 
structures of reduced dimensions (e.g., nanowires) with methodologies
developed in cluster physics, as well as highlighting certain nanowire
phenomena of mesoscopic origins and their analogies to clusters.

\section{The jellium model for metallic nanowires: Theoretical
method and results}

Consider a cylindrical jellium wire of length $L$, having a positive 
background with a circular cross section of radius $R \ll L$.
For simplicity, we restrict ourselves here to this symmetry of the wire
cross section. Variations in the shape of the nanowire cross section
serve to affect the degeneracies of the electronic spectrum 
\cite{sche,boga3} without
affecting our general conclusions. We also do not include here 
variations of the wire's shape along its axis. Adiabatic variation of
the wire's axial shape introduces a certain amount of smearing of the 
conductance steps through tunnelling, depending on the axial radius of
curvature of the wire. \cite{sche,boga2,boga3} 
Both the cross-sectional and axial shape of the wire
can be included in our model in a rather straightforward manner.

The principal idea of the SCM is the separation of the
total LDA energy $E_T(R)$ as\cite{yann1,yann2,stru}
\begin{equation}
E_T(R) = 
\widetilde{E}(R) 
+ \Delta E_{\text{sh}} (R)~,
\label{etscm}
\end{equation}
where $\widetilde{E}(R)$ varies smoothly as a function of the system size, and
$\Delta E_{\text{sh}} (R)$ 
is an oscillatory term arising from the discrete 
quantized nature of the electronic levels. $\Delta E_{\text{sh}} (R)$ is 
usually called a shell correction in the nuclear \cite{stru}
and cluster \cite{yann1,yann2} literature; we
continue to use here the same terminology with the understanding that the 
electronic levels in the nanowire form subbands, which are the analog of
electronic shells in clusters, where furthermore the size of the system is 
usually given by specifying the number of atoms $N$. The SCM method,
which has been shown to yield results in excellent agreement with experiments
\cite{yann1,yann3,yann5}
and self-consistent LDA calculations 
\cite{yann1,yann2} for a number of cluster systems, is 
equivalent to a Harris functional
($E_{\text{harris}}$) 
approximation to 
the Kohn-Sham LDA with the input density obtained through variational
minimization of an extended Thomas-Fermi (ETF) energy functional,
$E_{\text{ETF}}[\widetilde{\rho}]$
(with the kinetic energy,
$T_{\text{ETF}}[\widetilde{\rho}]$,
given to 4th order gradients and the potential,
$V_{\text{ETF}}$,
including the Hartree repulsion and exchange-correlation and 
positive-background attractions as in LDA). 
The smooth contribution in
Eq. (\ref{etscm}) is identified with 
$E_{\text{ETF}}[\widetilde{\rho}]$. The optimized density
\cite{note1}
$\widetilde{\rho}$ at a given radius $R$ is obtained under the
normalization condition (charge neutrality)
$2 \pi \int \widetilde{\rho} (r) r dr=
\rho^{(+)}_L(R)$, where $\rho^{(+)}_L (R) = 3R^2/(4r_s^3)$ is the
linear positive background density. Using the optimized 
$\widetilde{\rho}$, one solves for the eigenvalues
$\widetilde{\epsilon}_i$ 
of the Hamiltonian 
$H=-(\hbar^2/2m) \nabla^2 +
V_{\text{ETF}}[\widetilde{\rho}]$,
and the shell correction is given by
\begin{eqnarray}
\Delta E_{\text{sh}} & \equiv &
E_{\text{harris}}[\widetilde{\rho}] -
E_{\text{ETF}}[\widetilde{\rho}] \nonumber \\
 & = & \sum_{i=1}^{\text{occ}}
\widetilde{\epsilon}_i -
\int \widetilde{\rho} ({\bf r}) 
V_{\text{ETF}}[\widetilde{\rho} ({\bf r})] d{\bf r} -
T_{\text{ETF}}[\widetilde{\rho}]~,
\label{shcor}
\end{eqnarray}
where the summation extends over occupied levels.
here the dependence of all quantities on the pertinent size variable
(i.e., the radius of the wire $R$) is not shown explicitly. Additionally, the
index $i$ can be both discrete and continuous, and in the latter case the
summation is replaced by an integral.
 
Following the above procedure with a uniform background density of sodium
($r_s=4$ a.u.), a typical potential $V_{\text{ETF}}(r)$ for $R=12.7$ a.u., 
where $r$ is the radial
coordinate in the transverse plane, is shown in Fig.\ 1, along with the
transverse eigenvalues 
$\widetilde{\epsilon}_{nm}$
and the Fermi level; to simplify the calculations of the electronic
spectrum, we have assumed (as noted above) $R \ll L$, which allows us to
express the subband electronic spectrum as 
\begin{equation}
\widetilde{\epsilon}_{nm} (k_z;R) =
\widetilde{\epsilon}_{nm}(R) + \frac{\hbar^2 k_z^2}{2m}~,
\label{eigval}
\end{equation}
where $k_z$ is the electron wave number along the axis of the wire ($z$).

As indicated earlier,
taking the wire to be charge neutral, the electronic linear density,
$\rho_L^{\text{(--)}}$($R$), must equal the linear positive background density,
$\rho^{(+)}_L (R)$. 
The chemical potential (at $T=0$ the Fermi
energy $\epsilon_F$) for a wire of radius $R$
is determined by setting the expression for the
electronic linear density derived from the subband spectra
equal to $\rho^{(+)}_L (R)$, i.e.,
\begin{equation}
\frac{2}{\pi} 
\sum_{n,m}^{\text{occ}} \sqrt{ \frac{2m}{\hbar^2} [\epsilon_F (R)-
\widetilde{\epsilon}_{nm} (R)] } =
\rho^{(+)}_L (R)~,
\label{linden}
\end{equation}
where the factor of 2 on the left is due to the spin degeneracy. 
The summand defines the Fermi wave vector for each subband,
$k_{F,nm}$. The resulting
variation of $\epsilon_F (R)$ versus $R$ is displayed in Fig.\ 2(a), showing
cusps for values of the radius where a new subband drops below the Fermi
level as $R$ increases (or conversely as a subband moves above the Fermi
level as $R$ decreases upon elongation of the wire).
Using the Landauer expression for the conductance $G$ in the limit of
no mode mixing and assuming unit transmission coefficients, 
$G(R) = g_0 \sum_{n,m} \Theta [\epsilon_F(R)-\widetilde{\epsilon}_{nm}(R)]$, 
where $\Theta$ is the
Heaviside step function. The conductance of the nanowire, shown in Fig.\ 2(b),
exhibits quantized step-wise behavior, with the step-rises coinciding with the
locations of the cusps in $\epsilon_F (R)$, and the height sequence of the 
steps is 1$g_0$, 2$g_0$, 2$g_0$, 1$g_0$, ..., reflecting the circular
symmetry of the cylindrical wires' cross sections,
\cite{boga1} as observed for sodium nanowires. \cite{krans}  Solving for
$\epsilon_F (R)$ [see Eq.\ (\ref{linden})], the expression for the sum on the
right-hand-side of Eq.\ (\ref{shcor}) can be written as
\begin{eqnarray}
&& \sum_i^{\text{occ}} \widetilde{\epsilon}_i  =  
\frac{2}{\pi} \sum_{n,m}^{\text{occ}} 
\int_0^{k_{F,nm}} dk_z \widetilde{\epsilon}_{nm}(k_z;R)  =  
\nonumber \\
&& \frac{2}{3\pi} 
\sum_{n,m}^{\text{occ}}
[\epsilon_F(R) + 2 \widetilde{\epsilon}_{nm}(R)]
\sqrt{ \frac{2m}{\hbar^2} [\epsilon_F(R) - 
\widetilde{\epsilon}_{nm}(R)] }~, 
\label{sumi}
\end{eqnarray}
which allows one to evaluate 
$\Delta E_{\text{sh}}$ [Eq.\ (\ref{shcor})] for each wire radius $R$. Since the
expression in Eq.\ (\ref{sumi}) gives the energy per unit length, we also 
calculate $E_{\text{ETF}}$, $T_{\text{ETF}}$, and the volume integral in
the second line of Eq.\ (\ref{shcor}) for cylindrical volumes of unit height.
To convert to energies per unit volume [denoted as $\varepsilon_T (R)$,
$\widetilde{\varepsilon}(R)$, and $\Delta \varepsilon_{\text{sh}} (R)$] all
energies are further divided by the wire's cross-sectional area, $\pi R^2$.
The smooth contribution and the shell correction to the wire's energy are
shown respectively in Fig.\ 3(a) and Fig.\ 3(b). The smooth contribution
decreases slowly towards the bulk value ($-$2.25 eV per atom \cite{yann2}). 
On the other
hand, the shell corrections are much smaller in magnitude and exhibit an
oscillatory behavior. This oscillatory behavior remains 
visible in the total energy
[Fig.\ 3(c)] with the local energy minima occurring for values $R_{\text{min}}$
corresponding to conductance plateaus. The sequence of $R_{\text{min}}$ values
defines the MWC's, that is a sequence of wire configurations of enhanced
stability.

From the expressions for the total energy of the wire [i.e., 
$\Omega \varepsilon_T (R)$, 
where $\Omega = \pi R^2 L$ is the volume of the wire]
and the smooth and shell (subband) contributions to it, we can calculate the
``elongation force'' (EF),
\begin{eqnarray}
F_T (R) & = & -\frac{d[\Omega \varepsilon_T (R)]}{dL} 
= - \Omega \left\{ \frac{d \widetilde{\varepsilon} (R)}{dL} + 
\frac{ d [\Delta \varepsilon_{\text{sh}} (R) ]}{dL} 
\right\} \nonumber\\
 & \equiv &   \widetilde{F}(R) + \Delta F_{\text{sh}}(R)~.
\label{force}
\end{eqnarray}
Using the volume conservation, i.e., $d(\pi R^2 L)=0$, these forces can be
written as 
$F_T(R)= (\pi R^3/2) d\varepsilon_T (R)/dR$,
$\widetilde{F}(R)= (\pi R^3/2) d\widetilde{\varepsilon} (R)/dR$, and
$\Delta F_{\text{sh}} (R)= 
(\pi R^3/2) d [\Delta \varepsilon_{\text{sh}} (R)]/dR$.
$\widetilde{F}(R)$ and $F_T(R)$ are 
shown in Fig.\ 3(d,e). The 
oscillations in the force resulting from the shell-correction
contributions dominate. In all cases, the radii corresponding
to zeroes of the force situated on the left of the force maxima coincide with
the minima in the potential energy curve of the wire, corresponding to the
MWC's.  Consequently, these forces may be interpreted as guiding the 
self-evolution of the wire toward the MWC's.
Also, all the local maxima in the force occur at the locations of
step-rises in the conductance [reproduced in Fig.\ 3(f)], 
signifying the sequential decrease in the number of
subbands below the Fermi level (conducting channels) as the wire narrows 
(i.e., as it is being elongated). Finally the magnitude of the total forces
is comparable to the measured ones (i.e., in the nN range).

\section{Conclusions and Discussion}

We investigated energetics, conductance, and mesoscopic forces in a jellium 
modelled nanowire (sodium) using the local-density-functional-based shell 
correction method. 
The results shown above, particularly, the oscillations in the total energy
of the wire as a function of its radius (and consequently the oscillations
in the EF), the corresponding discrete sequence of magic wire configurations, 
and the
direct correlation between these oscillations and the step-wise quantized 
conductance of the nanowires, originate from quantization of the
electronic states (i.e., formation of subbands) due to the reduced lateral
(transverse) dimension of the nanowires. In fact such oscillatory behavior,
as well as the appearance of ``magic numbers'' and ``magic configurations'' of
enhanced stability, are a general characteristic of
finite-size fermionic systems and are in direct analogy
with those found in simple-metal clusters
(as well as in $^3$He clusters \cite{yann5} and atomic nuclei\cite{stru}), 
where electronic shell effects on the
energetics \cite{heer,yann1,yann2,yann3} 
(and most recently shape dynamics \cite{yann4} 
of jellium modelled clusters driven by forces obtained from shell-corrected 
energetics) have been studied for over a decade. 

While these calculations provide a useful and instructive framework, 
we remark that they are not a substitute for theories where the atomistic
nature and specific atomic arrangements are included 
\cite{land1,land2,land3,land4,barn} in evaluation of the
energetics (and dynamics) of these systems (see in particular Refs. 
\onlinecite{land4,barn}, where 
first-principles molecular-dynamics simulations of electronic spectra,
geometrical structure, atomic dynamics, electronic transport and
fluctuations in sodium 
nanowires have been discussed). 

Indeed, the atomistic structural
characteristics of nanowires (including the occurrence of cluster-derived 
structures of particular geometries\cite{land4,barn}), 
which may be observed through the use of
high resolution microscopy, \cite{kizu}
influence the electronic spectrum and 
transport characteristics, as well as the energetics of nanowires and
their mechanical properties and response mechanisms. In particular,
the mechanical response of materials involves structural changes
through displacement and discrete rearrangenent of the atoms. The 
mechanisms, pathways, and rates of such structural transformations
are dependent on the arrangements and coordinations of atoms, the
magnitude of structural transformation barriers, and the local shape of the
wire, as well as possible dependency on the history of the material
and the conditions of the experiment (i.e., fast versus slow
extensions). Further evidence for the discrete atomistic nature of the 
structural transformations is provided by the shape of the force variations
(compare the calculated Fig.\ 3(b) in Ref.\ \onlinecite{land1} and
Fig.\ 3 in Ref.\ \onlinecite{land2} with the measurements shown in Figs.\ 1 
and 2 in Ref.\ \onlinecite{rubi}), and the interlayer spacing period of the 
force oscillations when the wire narrows. 
While such issues are not addressed by 
our model, the mesoscopic (in a sense universal) phenomena described
by it are of interest, and may guide further research in the area of 
finite-size systems in the nanoscale regime.

\acknowledgments
This research was supported by a grant from the U.S. Department of Energy
(Grant No. FG05-86ER45234) and the AFOSR.
Useful conversations with W.D. Luedtke, E.N. Bogachek, and R.N. Barnett
are greatfully acknowledged. Calculations were performed at the Georgia
Institute of Technology Center for Computational Materials Science.

\begin{figure}
\caption{Lower panel: The $V_{\text{ETF}}(r)$ potential for a sodium wire 
with a uniform jellium background of radius $R=12.7$ a.u., plotted versus
the transverse radial distance from the center of the wire, along with
the locations of the bottoms of the subbands (namely the transverse 
eigenvalues $\widetilde{\epsilon}_{nm}$; $n$ is the number of nodes in the
radial direction plus one, and $m$ is the azimuthal quantum number of the 
angular momentum). The Fermi level is denoted by a dashed line.
Top panel: The jellium background volume density (dashed line) and
the electronic volume density $\widetilde{\rho}(r)$ (solid line, exhibiting
a characteristic spillout) normalized to bulk values are shown.
}
\end{figure}
\begin{figure}
\caption{
Variation of the Fermi energy $\epsilon_F$ [shown in (a)] and of the
conductance $G$ (shown in (b) 
in units of $g_0=2 e^2/h$), plotted versus the radius R,
for a sodium nanowire. Note the coincidence of the cusps in $\epsilon_F$
with the step-rises of the conductance. The heights of the steps in $G$
reflect the subband degeneracies due to the circular shape of the wire's
cross section.
}
\end{figure}
\begin{figure}
\caption{
(a-c): The smooth (a) and shell-correction (b) contributions to the total
energy (c) per unit volume of the jellium-modelled sodium nanowire (in units of
$u \equiv 10^{-4}\;$eV/a.u.$^3$), 
plotted versus the radius of the wire (in a.u.). Note the
smaller magnitude of the shell corrections relative to the smooth
contribution. (d-e): The smooth contribution (d) to the total force and
the total force (e), plotted in units of nN versus the wire's radius.
In (e), the zeroes of the force to the left of the force maxima occur at
radii corresponding to the local minima of the energy of the wire (c).
In (f), we reproduce the conductance of the wire (in units of 
$g_0=2e^2/h$), plotted versus R.
Interestingly, calculations of the conductance for the MWC's (i.e.,
the wire radii corresponding to the locations of the step-rises)
through the Sharvin-Weyl formula, 
\protect\cite{garc,boga3} 
corrected for the finite height of the
confining potential 
\protect\cite{garc} 
(see lower panel of Fig.\ 1), namely
$G=g_0 (\pi S/\lambda_F^2 - \alpha P/\lambda_F$) where $S$ and $P$ are the
area and perimeter of the wire's cross section and $\lambda_F$ is the
Fermi wavelength ($\lambda_F=12.91$ a.u. for Na) with $\alpha=0.1$
(see Ref.\ 
\protect\onlinecite{garc}), 
yield results which approximate well the
conductance values (i.e., the values at the bottom of the step-rises)
shown in (f).
}
\end{figure}

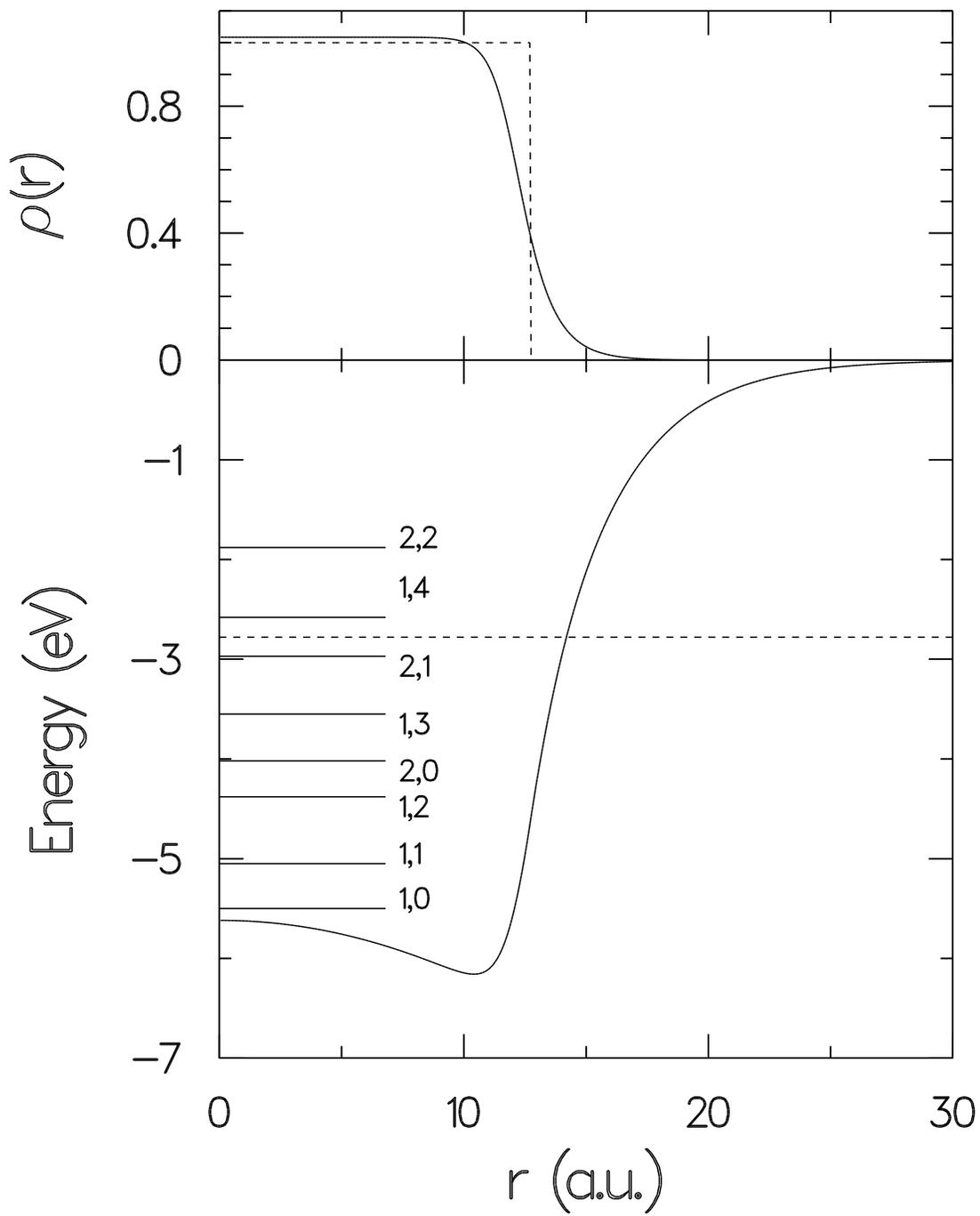
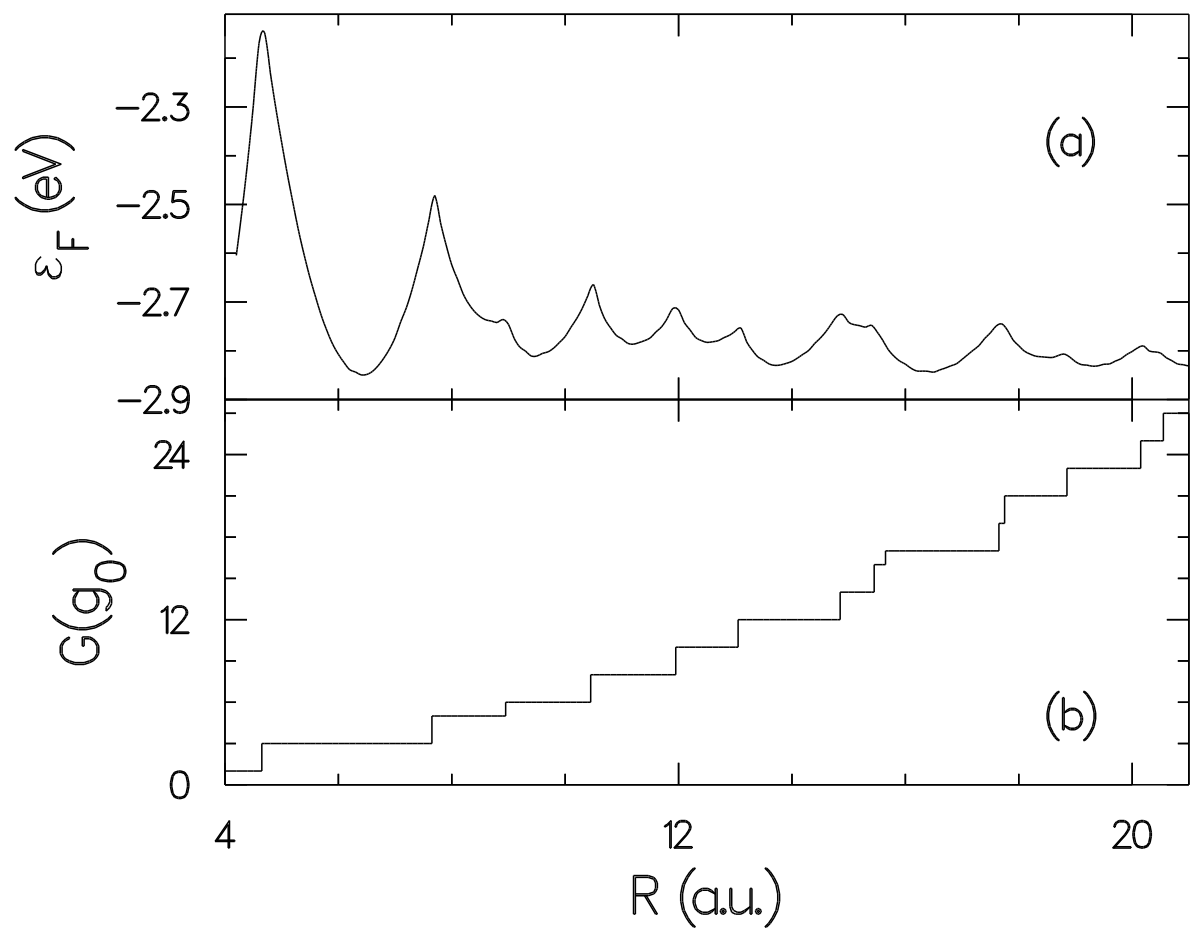
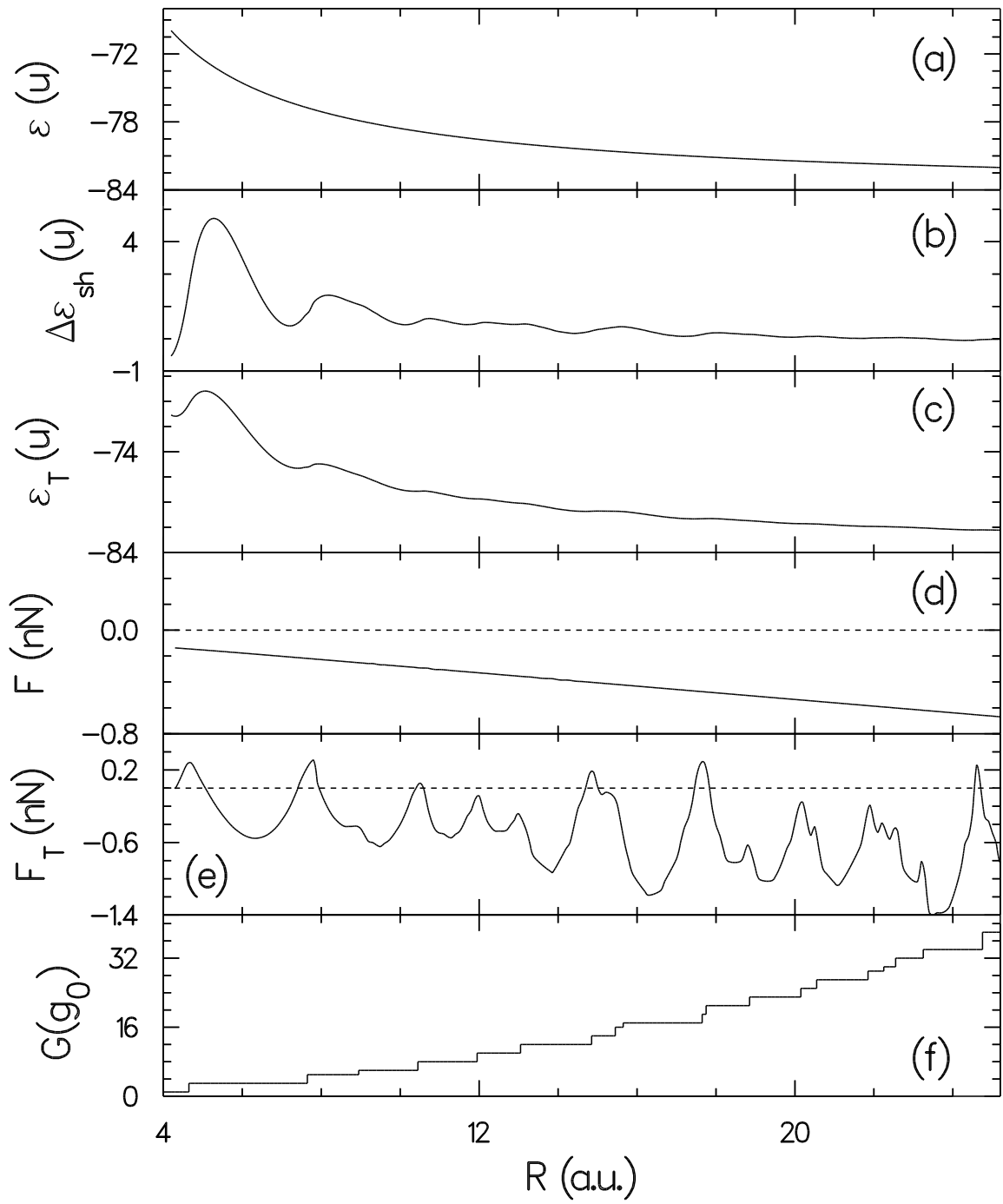
\end{document}